# Temperature dependence of relativistic valence band splitting induced by an altermagnetic phase transition

M. Hajlaoui[1], S.W. D'Souza[2], L. Šmejkal[3,4], D. Kriegner[3], G. Krizman[1], T. Zakusylo[1], N. Olszowska[5], O. Caha[6], J. Michalička[7], A. Marmodoro[2,3], K. Výborný[3], A. Ernst[8], M. Cinchetti[9], J. Minar[2], T. Jungwirth[3,10], and G. Springholz[1]

[1]Institute of Semiconductors and Solid-State Physics, Johannes Kepler University, 4040 Linz, Austria
[2]University of West Bohemia, New Technologies Research Center, Pilsen 30100, Czech Republic
[3]Institute of Physics, Czech Academy of Sciences, Cukrovarnická 10, 162 00 Praha 6, Czech Republic
[4]Institute of Physics, Johannes Gutenberg University Mainz, D-55099 Mainz, Germany
[5]National Synchrotron Radiation Centre SOLARIS, Jagiellonian University, Czerwone Maki 98, 30-392 Krakow, Poland
[6]Department of Condensed Matter Physics, Masaryk University, Kotlářská 267/2, Brno 61137, Czech Republic
[7]Central European Institute of Technology, Brno University of Technology, Purkyňova 123, Brno, 61200, Czech Republic
[8]Institute for Theoretical Physics, Johannes Kepler University, 4040 Linz, Austria
[9]Department of Physics, TU Dortmund University, 44227 Dortmund, Germany
[10]School of Physics and Astronomy, University of Nottingham, Nottingham NG7 2RD, United Kingdom

**Altermagnetic (AM) materials exhibit non-relativistic, momentum-dependent spin-split states, ushering in new opportunities for spin electronic devices. While the characteristics of spin-splitting have been documented within the framework of the non-relativistic spin group symmetry, there has been limited exploration of the inclusion of relativistic symmetry and its impact on the emergence of a novel spin-splitting in the band structure. This study delves into the intricate relativistic electronic structure of an AM material, α-MnTe. Employing temperature-dependent angle-resolved photoelectron spectroscopy across the AM phase transition, we elucidate the emergence of a relativistic valence band splitting concurrent with the establishment of magnetic order. This discovery is validated through disordered local moment calculations, modeling the influence of magnetic order on the electronic structure and confirming the magnetic origin of the observed splitting. The temperature-dependent splitting is ascribed to the advent of relativistic spin-splitting resulting from the strengthening of AM order in α-MnTe as the temperature decreases. This sheds light on a previously unexplored facet of this intriguing material.**

## 1. Introduction

Exploring the unique properties of electron spins and tailoring their spin splitting holds significant potential for the development of more efficient and versatile spin-electronic devices **[1][2][3][4][5]**. In conventional antiferromagnets with antiparallel magnetic order, symmetries usually prevent a spin splitting of bands **[6]**. For this reason, materials where those symmetries are broken have been considered **[7][8][9][10]**. This lead to the recent theoretical breakthrough that allowed to identify a new class of materials termed 'altermagnets' (AMs) **[11]** which are distinct from conventional (collinear) antiferromagnets and exhibit a momentum-dependent spin splitting despite the compensation of magnetic moments. The spin splitting in AMs can be of non-relativistic origin, displaying energy splitting orders of magnitude larger than those typically induced by relativistic spin-orbit coupling (SOC).

The direction of the spin polarization alternates throughout the Brillouin zone, changing sign upon crossing symmetry-enforced nodal surfaces where spin degeneracy in the non-relativistic electronic structure is maintained. Altermagnetic crystals can exhibit two, four, or six spin-degenerate nodal surfaces, corresponding to d-wave, g-wave, or i-wave type of the non-relativistic spin-polarization order, respectively **[6][11]**. Here, we focus on α-MnTe, identified as a g-wave altermagnet in the absence of SOC **[6][11]**. We delve into the implications of the relativistic SOC in the AM-MnTe in inducing additional relativistic spin splitting on these nodal surfaces.

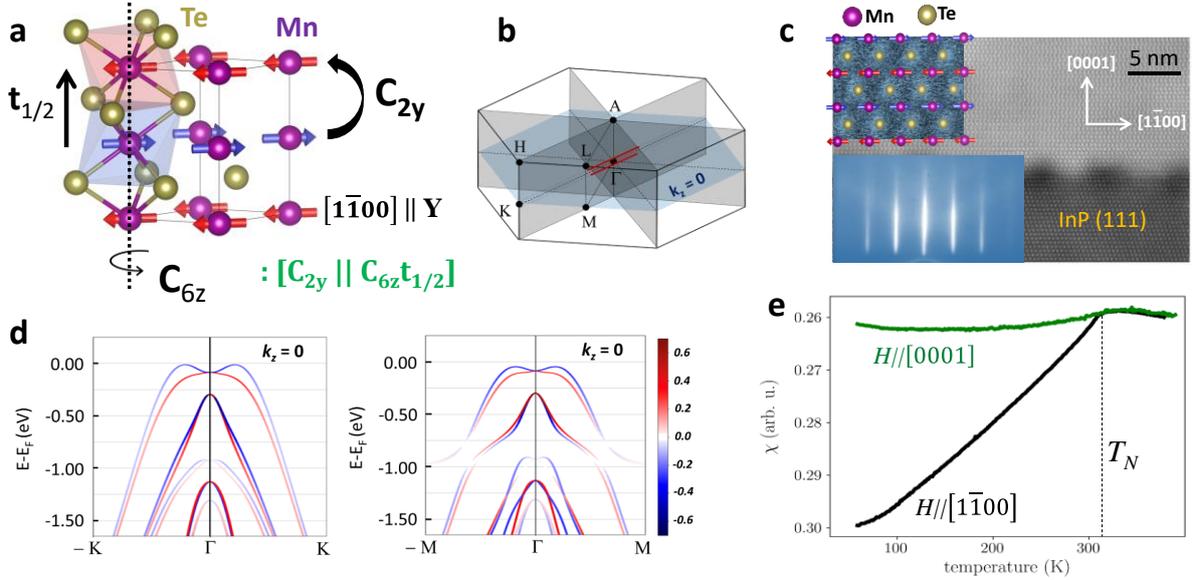

**Figure1: Properties of altermagnetic α-MnTe.** (a) AM structure of α-MnTe with opposing spin-sublattices connected by a non-symmorphic six-fold screw-axis rotation $C_6t_{1/2}$ combined with $C_2$ spin rotation: $[C_2||C_{6z}t_{1/2}]$. The Néel vector is oriented along the $[1\bar{1}00]$ direction. (b) Bulk Brillouin zone, displaying the four nodal degenerate surfaces (grey and blue planes). The two red lines delineate the cuts in k-space on the nodal surfaces (AHKΓ) where temperature-dependent ARPES is conducted at hν = 21 eV and 48 eV (see, Fig. 3). (c) High-angle annular dark-field scanning transmission electron microscopy (HAADF-STEM) image along the $[11\bar{2}0]$ direction. A corrugation of approximately five atomic layers is observed at the MnTe/InP(111) interface, which is attributed to the high-temperature annealing necessary to remove the native oxide of the InP surface. A zoomed view of the MnTe epilayer is displayed in the upper left corner, while the lower left inset presents in situ RHEED patterns recorded at the MnTe (0001) surface during MBE growth. (d) The electronic structure of AM-MnTe along the Γ−K and Γ−M directions with spin-orbit coupling, highlighting the spin-polarized states at the $k_z$ = 0 plane, with red and blue colors representing spin up and spin down, respectively. (e) Temperature-dependent susceptibility measured under different magnetic field directions.

α−MnTe crystallizes in the hexagonal NiAs- structure with a space group symmetry of P6₃/mmc (see Fig. 1a). In the magnetically ordered phase, the Mn moments align parallel within each c-plane but stack antiparallel along the c-axis, resulting in a collinear magnetically compensated phase. Without SOC, the spin sublattices are connected by a non-symmorphic six-fold screw-axis rotation $C_6t_{1/2}$ combined with $C_2$ spin rotation: $[C_2||C_{6z}t_{1/2}]$. The resulting non-relativistic spin-group symmetry (²6/²m²m¹m) reveals a strong time reversal symmetry breaking and alternating spin splitting in the band structure **[6][11][12]**. Consequently, α−MnTe falls into the g-wave AM category, characterized by three spin-degenerate nodal planes crossing the Γ−point and parallel to the c-axis (grey planes in Fig 1.b), and one more spin-degenerate nodal plane crossing the Γ−point orthogonal to the c-axis (the $k_z$ = 0 plane; the blue plane in Fig. 1b). When SOC interaction is introduced, the relativistic magnetic point group varies depending on the Néel vector orientation [12]. By inclusion of SOC in the AM symmetry, the spin degeneracy within these nodal planes is lifted. This effect is illustrated in Fig. 1d for the case of $k_z$ = 0 plane, along the $\Gamma-K$ and $\Gamma-M$ directions, depicting the spin-polarized band structure, with red and blue colors representing spin up and spin down, respectively [13].

The experimental exploration of AM properties of $\alpha-$MnTe has started with the observation of the spontaneous anomalous Hall effect (AHE) [12][14], serving as a signature rather than direct evidence for AM. Angle-resolved photoemission spectroscopy (ARPES) thus, plays a pivotal role in investigating the altermagnetism of $\alpha-$MnTe [13][15][16]. Soft X-ray ARPES and Spin-ARPES have been recently employed to measure the bulk and spin polarized electronic structure, providing the first experimental evidence of AM lifting of Kramer's spin degeneracy [13]. In addition, ARPES along *k*-space directions not aligned with nodal surfaces was used to measure the energy splitting induced by the AM phase transition [15] and micro-focused ARPES on a single crystal provided insights into the anisotropic spin-split band structure of the AM phase [16]. The application of ARPES also identified the AM properties in other prototypical materials, such as $RuO_2$ [17] and CrSb [18].

In the present work, we present temperature-dependent UV-ARPES measurements spanning over the whole magnetic phase transition, with particular attention directed towards states on the nodal surfaces. The results unveil a temperature-dependent relativistic spin splitting that diminishes gradually as the system approaches the paramagnetic (PM) phase. Through utilization of disordered local moment calculations to model the PM-AM transition, we demonstrate the magnetic origin of the AM valence band splitting. The microscopic rationale for this magnetism-induced splitting lies in the establishment of the relativistic AM symmetry as the temperature decreases below the Néel temperature. This sheds light on an important unexplored facet of this intriguing class of materials.

## 2. Results and Discussion

For our investigations, single crystalline $\alpha-$MnTe films were grown on InP (111) substrates using molecular beam epitaxy, as detailed in the methods section. Under optimized conditions, we obtained high-quality epitaxial layers with hexagonal NiAs-type structure in the (0001) orientation and excellent surface properties. The characterization of the film was carried out using transmission electron microscopy (TEM) and reflection high-energy electron diffraction (RHEED), as shown in Fig. 1c. X-ray diffraction and atomic force microscopy were also employed, yielding consistent results with our previous works [19][20][21][22]. For further insights on the magnetic transition, we conducted temperature-dependent susceptibility measurements under various magnetic field directions, revealing an AM phase transition at $T_N$ = 307 K (see Fig. 1e). For UV-ARPES investigations, the films were transferred from our molecular beam epitaxy system to the URANOS beamline at the SOLARIS synchrotron facility without breaking the ultra-high vacuum conditions (< $10^{-10}$ mbar).

### 2.1 The relativistic altermagnetic electronic structure

To reveal the electronic band structure of the AM phase, we first focus on the ARPES photon energy dependence at *T* = 40 K to elucidate the dispersion of the band structure along the $k_z$ – direction and thereby, determine the positions of the high symmetry points $\Gamma$ and $A$ as a function of photon energy in the range of 19 eV to 65 eV. Pure surface states typically do not disperse with photon energy, whereas surface resonances and bulk projected bands usually significantly disperse along the $k_z$ direction. Figure 2a presents integrated photon energy-dependent ARPES spectra within an energy window of 50 meV, centered at a binding energy of 150 meV below the Fermi level. This exclusive focus reveals the dispersion of states corresponding to band $B_1$. The results indicate a distinct $k_z$-dispersion of $B_1$ across different photon energies. A periodic pattern in the dispersion is noticeable, particularly around hv = 22 and 46 eV, where a broader $k_{||}$ momentum dispersion is observed. Conversely, at hv = 36 and 64 eV, the map illustrates dispersion within a narrower $k_{||}$ range. The $\Gamma$ and $A$ points have been attributed to these photon energies, respectively. This is supported by Fig. 2b, showing the positions of $\Gamma$ and $A$ in $k_z$ values, assuming a parabolic final state with an inner potential $V_0$ = 11 eV. The same inner potential was also used to calculate the $k_z$-dependence of the soft-X-ray ARPES data, revealing a consistent dispersion as in one-step photoemission calculations **[13]**. This observation suggests that $B_1$ may exhibit characteristic features indicative of its bulk nature. We would like to note that because the data in Fig. 2a shows the $k_z$ dispersion at 100 - 200 meV below the Fermi level, it does not indicate the position of the valence band maximum, as discussed in Ref. **[23]**.

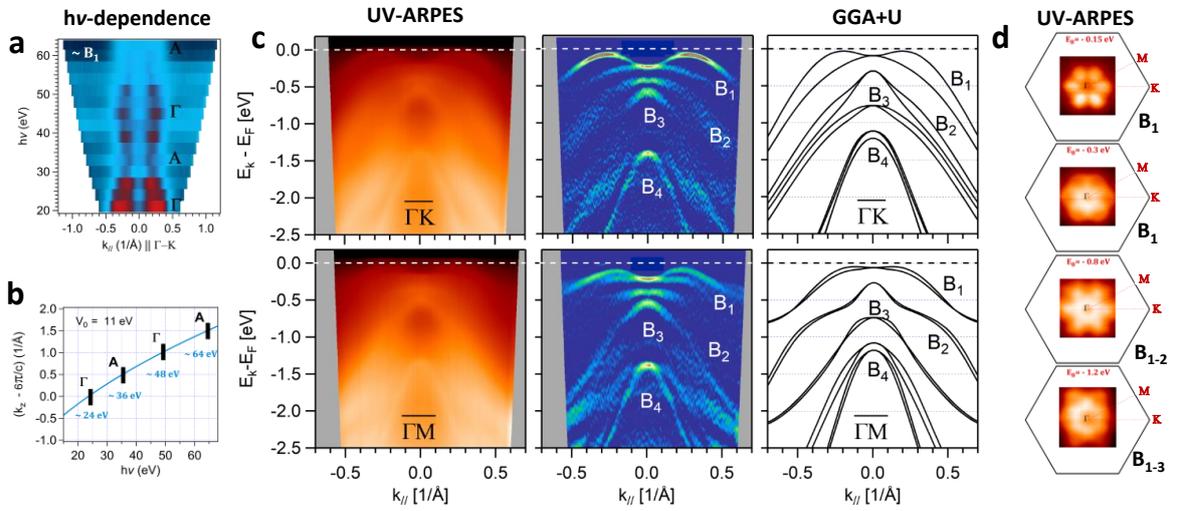

**Figure 2: The relativistic electronic structure of α-MnTe:** (a) Photon energy dependence of UV-ARPES integrated over the bulk $B_1$ band near the Fermi level at T = 40 K (AM-phase). Here, the red color corresponds to the highest intensity. (b) The $k_z$ values as a function of photon energy at normal emission and near the Fermi level, assuming a parabolic final state of bands and an inner potential of $V_0$ = 11 eV. The positions of $\Gamma$ and $A$ are indicated. (c) UV-ARPES $E(k_{||})$ band maps (left) and their $2^{nd}$ derivative in blue (middle) recorded at hv = 21 eV at 40 K along the $\Gamma-K$ (top) and $\Gamma-M$ directions (bottom). The corresponding bulk GGA+U calculations are shown on the right, showcasing the emergence of three groups of bulk bands $B_1$, $B_2$, $B_3$ each consisting of two spin split bands, and the group $B_4$ consisting of four splitted bands. (d) Constant energy cuts of the UV-ARPES data mapped to the ($k_x$, $k_y$) plane at different energies of -0.15, -0.3, -0.8 and -1.2 eV below the Fermi level.

For further analysis, we show in Fig. 2c the $E(k_{||})$ ARPES maps recorded at 21 eV close to the $\Gamma$-point, along the two high-symmetry $k_{||}$-directions $\Gamma-K$ (top) and $\Gamma-M$ (bottom), respectively. For better visualization, the second derivative of these maps is shown in the middle panels to highlight the individual band dispersions. Evidently, the band $B_1$ splits up in two branches away from the $\Gamma$-point with a nearly quadratic increase of the splitting with increasing $k_{||}$, which is more pronounced in the $\Gamma-K$ compared to the $\Gamma-M$ direction. This unique feature is attributed to the spin splitting of the AM bands shown in Fig. 1d, and it perfectly agrees with the bulk GGA+U calculations including SOC shown on the right-hand side of Fig. 2c. The importance of including SOC is shown in the supplementary material Note 1, where present the comparison of the ARPES data with the GGA+U without SOC, clearly evidencing the spin splitting due the relativistic effect in the electronic structure.

The temperature-dependent measurements presented below provide additional clarity, reinforcing that the observed energy splitting is a result of spin splitting induced by the AM order. The comparison with the GGA+U calculation reveals that in ARPES we consistently observe the energy-momentum dispersions for all bands $B_1$, $B_2$, $B_3$, and $B_4$, confirming their dominant bulk character. This is elucidated further in the next section through a comparison of the temperature dependence of these bands and disordered local moment (DLM) calculations that model the PM-AM transition.

It is important to consider the effect of SOC in the presence of the magnetic order. In fact, as outlined in the supplementary information (Note2) and detailed in Refs. **[23] [24]**, the combination of an in-plane Néel vector and SOC interaction breaks the in-plane six-fold symmetry of the electronic structure **[25]**. In real materials, however, the magnetic structure will form domains with different orientations of the Néel vector along one of the six equivalent in-plane easy axes **[19][20][26]**. The size of these domains typically varies strongly as a function of temperature and sample growth conditions, but is expected to be well below the spot size of our UV-ARPES measurements (> 100 μm). As a result, in our measurements we average the band structure over multiple domains. This averages out the anisotropy as is demonstrated in Fig. 2d by the constant energy cuts of the ARPES data, which display a perfect hexagonal symmetry of the band maps for all energies below the Fermi level, with no discernible evidence of anisotropy or the presence of two-fold symmetry. Thus, the present ARPES data corresponds to the domain averaged band structure of the system. As shown in the supplemental material (Note2 and Fig. S3), fortuitously, the band structure varies only slightly from one domain to another, so that the main features of the bands – namely their splitting, remains well visible in our experiments even when multiple domains are superimposed. As a result, the large relativistic AM splitting of the band $B_1$ is well visible in our ARPES maps even after the averaging process.

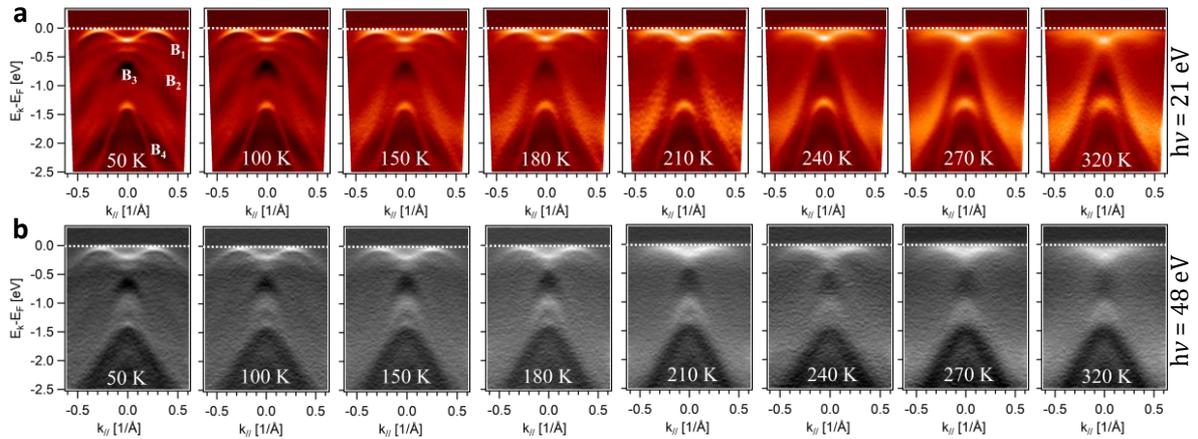

**Figure 3: Temperature-dependent spin splitting across the AM–PM transition.** Temperature dependent UV-ARPES maps from 50 to 320 K recorded parallel to the Γ−K direction. Data shown in (a) in red was recorded at a photon energy hν = 21 eV, data at hν = 48 eV (b) is shown in grey scale. The cut-lines in k-space corresponding to these two-photon energies are highlighted as two red-lines in Fig. 1b.

## 2.2 Spin splitting induced by the altermagnetic phase transition

To elucidate the influence of AM order on the lifting of spin degeneracy during the magnetic phase transition, we systematically studied temperature-dependent ARPES from 50 K (below $T_N$) to 320 K (above $T_N$). Figure 3a presents the resulting ARPES data (first derivative) acquired at *hν* = 21 eV along Γ−K, revealing a strong temperature evolution of the band structure. The same was measured at *hν* = 48 eV as depicted in Fig. 3b, where the corresponding $k_z$ value lies also in close proximity to Γ (refer to Fig. 2a and 2b). At both 21 eV and 48 eV the ARPES maps are parallel to the Γ−K direction, as indicated by the respective red cut lines in Fig. 1b, capturing the electronic structure on the same nodal surface (AHKΓ, see Fig. 1b). The complete raw data and its derivatives are shown in the supplementary material (Note3).

Analyzing the ARPES images for *hv* = 21 eV (Fig. 3a), we clearly find that with increasing temperature the splitting of the $B_1$, $B_2$ and $B_3$ bands gradually diminishes, whereas from 270 K the splitting apparently disappears. Exactly the same is also observed for the spectra at *hv* = 48 eV (Fig. 3b). Notably, the drastic change of the splitting is most prominent for bands $B_1$, $B_2$, and $B_3$, while band $B_4$ remains more or less unaffected. Hence, our ARPES measurements at photon energies of 21 eV and 48 eV demonstrate a temperature-dependent valence band splitting on the nodal surface that is clearly induced by the altermagnetism of the system.

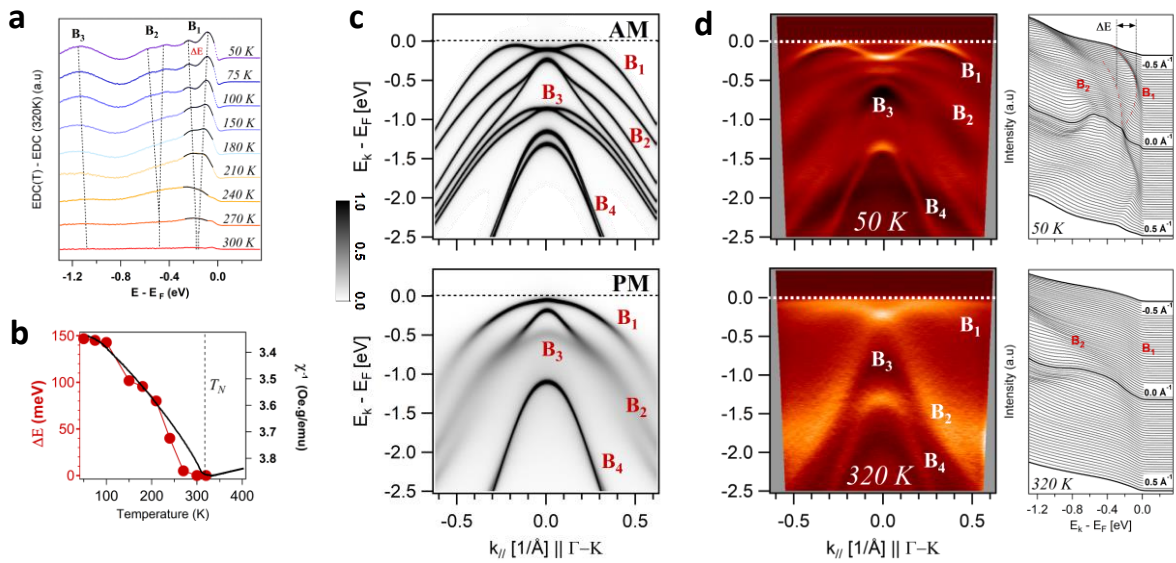

**Figure 4: Temperature-dependent splitting: ARPES versus DLM-CPA calculations.** (a) Momentum integrated difference ΔEDC (T <320K) ARPES profiles minus the EDC at 320K, i.e., ΔEDC(T) = EDC(T) – EDC (320K) as a function of temperature. (b) Derived splitting ΔE of $B_1$ (red symbols) plotted as a function of temperature, alongside with the temperature-dependent inverse susceptibility represented by the black line. (c) DLM-CPA calculations of the AM (top) and PM (bottom) phases, revealing the pronounced difference in the band structure and spin splitting in the AM and PM states. (d) Corresponding UV-ARPES maps at 50 K and 320 K below and above $T_N$ along the ΓK direction (1$^{st}$ Derivative, hv = 21 eV). The raw data of EDCs at different momenta are shown on the right.

The detailed analysis of the splitting evolution is presented in Fig. 4a, illustrating the pronounced change in the momentum-integrated energy dispersion curves (EDC) presented as ΔEDC(T) = EDC(T<320K) – EDC (320K), clearly demonstrating the strong increase of the splitting when the temperature is lowered from 270 to 50 K. In particular, the difference spectra clearly reveal the transition from the large splitting of bands $B_1$ and $B_2$ of the low-temperature AM phase to the merged bands in the high temperature PM phase. Simultaneously, the intensity of the $B_3$ band increases and it shifts to high binding energy. To evaluate the values of temperature-dependent splitting, we focus on the splitting of band $B_1$. We accomplish this by fitting the spectra of ΔEDC(T) with two Gaussian curves near these states (depicted as black curves in Fig. 4a). The results are presented in Fig. 4b, along with the temperature-dependent inverse susceptibility from Fig. 1e. Evidently, the splitting reaches a maximum value of approximately 150 meV at 50 K and gradually decreases until it becomes nearly indiscernible at 270 K. Both the splitting as well as the susceptibility exhibit a similar evolution with temperature, confirming the magnetic origin of the PM-AM transition in α−MnTe.

To further corroborate the observed valence band splitting, we compare the low and high temperature ARPES spectra to band structure calculation using the disordered local moment (DLM) approach in conjunction with the coherent potential approximation (CPA) to model the electronic structure of both the disordered PM and ordered AM phases (See Methods section). The results are displayed in Fig. 4c and compared to the ARPES spectra shown in Fig. 4d. The calculated band structure shows a remarkably good agreement with the ARPES data both for high temperature and low temperature phases, evidencing clearly that the observed band structure changes are indeed due to the PM-AM phase transition. Specifically, the splitting of bands $B_1$, $B_2$, and $B_3$ is consistently observed in both DLM-CPA and low temperature ARPES (50 K) but is absent for the high temperature PM phase, while the $B_4$ band remains nearly unaffected both in theory and experiments. This validates the bulk nature of these bands and affirms that the observed valence band splitting is of magnetic origin. Thus, the influence of the magnetic order on the energy splitting of the electronic structure is well-described by the DLM-CPA model, supporting our observation with ARPES.

The microscopic physics underlying the observed temperature-dependent valence band splitting can be attributed to the emergence of relativistic spin splitting resulting from the strengthening of AM symmetry as temperature decreases. As outlined in the introduction, the altermagnetism of α−MnTe implies that in the absence of SOC, the spin degeneracy is preserved on the nodal surfaces. Simultaneously, α−MnTe is centrosymmetric, meaning that SOC alone cannot induce spin-splitting states. Indeed, our calculations in Fig. 1d directly confirm that the unique combination of both SOC and AM order together leads to the emergence of a relativistic spin splitting on the nodal surfaces, emphasizing that the spin splitting on the nodal surfaces is fundamentally relativistic and requires the presence of AM order. Our ARPES spectra, acquired at 21 eV and 48 eV, capture the states on the nodal surface (cf. Fig. 1b). Moreover, our data showcases the strong increase in energy splitting of these states within the electronic structure as the temperature decreases. This observation is attributed to the presence and strengthening of the AM order, a conclusion supported by DLM-CPA calculations. The temperature-dependent ARPES spectra, thus, effectively reveal the impact of altermagnetism-induced spin splitting on the electronic structure manifested through emergent splitting of the valence bands. Thus, it occurs exclusively in the AM phase.

It's noteworthy that the spin-polarized band structure calculations were made under the assumption that the material exists in a single domain. Averaging over the multiple domains, as we do in our experiments, cancels out the spin polarization unless an unequal domain population is present by growth **[13]** or applying external magnetic fields. Thus, although our domain-averaged ARPES spectra do not yield information on the spin polarization, yet they effectively unveil the energy splitting of the electronic structure caused by the altermagnetic phase, which is absent in the PM phase. This is revealed in detail in the supplementary material (Note2 and Fig. S3), where it is shown that the energy splitting in the domain averaged band structure indeed retains the prominent band splitting even if the band structure of multiple domains is superimposed. Consequently, the large splitting of $B_1$ emerging due to the formation of the AM order is well resolved in our experiments. Thus, our observation of the temperature-dependent valence band splitting across the magnetic transition is a clear hallmark of the altermagnetic state of α-MnTe.

## 3. Conclusions

Through angle-resolved photoemission spectroscopy and density functional theory, we have elucidated the electronic structure of altermagnetic α–MnTe in both its low-temperature altermagnetic and high-temperature paramagnetic phases. In the altermagnetic phase, we identified a valence band splitting as large as 150 meV on the nodal surface that gradually diminishes as the temperature increases towards the Néel temperature, and ultimately vanishes above in the paramagnetic phase. This discovery strongly supports the notion that the valence band splitting in altermagnets is intricately linked to the unique momentum dependent spin splitting caused by the altermagnetic order, as confirmed by our disordered local moment calculations. The presence of a temperature-dependent valence band splitting on the nodal surface unequivocally reveals the crucial role of relativistic altermagnetic symmetry in inducing relativistic spin splitting in α–MnTe that stands for a prototypical altermagnetic material. Thus, our work not only expands our understanding of altermagnetic phenomena but also opens new avenues for the development of advanced materials with unique electronic properties. As we delve deeper into the intricacies of altermagnetic systems, we anticipate that further discoveries will drive advancements that have profound implications for both fundamental science and technological applications.

## 4. Methods Section

***Growth and characterization***. Single crystalline α-MnTe films were grown by molecular beam epitaxy on InP (111) A substrates using elemental Mn and Te sources and substrate temperatures in the range of 370 to 450°C. The structure of the films was determined using X-ray diffraction (XRD) and transmission electron microscopy (STEM) at room temperature. The surfaces were assessed using reflection high-energy electron diffraction (RHEED) and atomic force microscopy (AFM), while the magnetic transition was characterized through superconducting quantum interference device (SQUID) measurements.

***Angle-resolved Photoemission spectroscopy***. ARPES measurements were performed on the high-resolution URANOS beamline at the SOLARIS synchrotron in Krakow, Poland. For this purpose, the samples were transferred from the MBE to the synchrotron under UHV conditions using a battery-operated Ferrovac vacuum suitcase. In these experiments, the photon energy was varied within the range of 17 eV to 70 eV (p-polarized) and the sample temperature between 320 K and 40 K. The photoelectrons kinetic energy and their emission angles are measured by a VG Scienta DA30L electron spectrometer with energy and angular resolution better than 3 meV and 0.1°, respectively.

***Density functional calculation + U (DFT+U)***. To compare with ARPES spectra, self-consistent electronic structure calculations were performed for the magnetically ordered phase of MnTe (P63/mmc space group) using the lattice parameters a = 4.15 Å and c = 6.71 Å. These calculations utilized projector augmented wave pseudopotentials (PAW) **[27]** and adopting the implementation of the VASP electronic structure code **[28]**. We chose the exchange-correlation functional generalized gradient approximation (GGA) according to Perdew, Burke, and Ernzerhof (PBE) **[29]**. BZ integral were sampled on a Γ-centered 9 × 9 × 4 mesh, and a consistent energy cutoff of 400 eV was applied in all calculations. The screened on-site Coulomb interaction U and exchange interaction J for Mn were set to 4.80 eV and 0.80 eV, respectively, based on previous literature estimates **[30]**. In these calculations, spin-orbit coupling was included on-top to the Schrödinger Hamiltonian and the direction of the Néel vector was set along [1$\bar{1}$00] direction, in accordance with experimental evidence from neutron diffraction and magneto-transport measurements **[20]**. As detailed in the supplementary materials (Note2 and Fig. S3), it was observed that the band structure varies only slightly from one domain to another, enabling the main features of the bands to remain clearly visible in our experiments. In this case, it is assumed that the electronic structure along the ΓK and ΓM directions is equivalent to that along the ΓK$_3$ and ΓM$_2$ directions, respectively.

**Disordered local moment (DLM) combined with the coherent potential approximation (CPA).** Self-consistent calculations were carried out using the fully relativistic Korringa-Kohn-Rostoker (SPRKKR) Green's function method in the atomic sphere approximation (ASA), within the rotationally invariant GGA+U scheme as implemented in the SPRKKR formalism **[31][32].** The screened on-site Coulomb interaction U and exchange interaction J of Mn were set to 4.80 eV and 0.80 eV respectively, which were estimated based on the previous literature **[30]**. The angular momentum expansion up to lmax = 4 has been used for each atom on a 28 x 28 x 15 k-point grid. To simulate the paramagnetic phase, we employed the Disordered Local Moment (DLM) approach, assigning an equal 50% probability to either sign for the magnetic orientation of antiparallel Mn atoms at T ≥ T$_N$. The total energy convergence has been set to 10−5 Ry. The Lloyd's formula has been employed for accurate determination of the Fermi level **[33] [34] [35].**

## *Supporting Information*

Supporting Information is available from the Wiley Online Library or from the author

## *Acknowledgements*

M.H. and G.S. acknowledge the support by the Austrian Science Fund Grants P30960-N27 and I-4493-N. L.S. acknowledges support from JGU TopDyn initiative. L.S. acknowledges Deutsche Forschungsgemeinschaft (DFG, German Research Foundation) —TRR 288 – 422213477 (Project A09 and B05). S.W.D, O.C, and J.M thanks QM4ST project financed by the Ministry of Education of Czech Republic Grant No. CZ.02.01.01/00/22_008/0004572, co-funded by the ERD. D.K. acknowledges the support from the Czech Academy of Sciences (project No. LQ100102201) and Czech Science Foundation Grant No. 22-22000M. J.M acknowledge CzechNanoLab Research Infrastructure supported by MEYS CR (LM2023051). A.M. acknowledges the Czech Science Foundation Grant No. 23-04746S. M.C acknowledges financial support by the Deutsche Forschungsgemeinschaft through the International Collaborative Research Centre 160 (Projects No. B8 and No. Z4) and by the European Union's Horizon 2020 Research and Innovation Programme under Project SINFONIA, Grant 964396. The ARPES setup was developed under the provision of the Polish Ministry and Higher Education project Support for research and development with the use of research infra-structure of the National Synchrotron Radiation Centre "SOLARIS" under contract nr 1/SOL/2021/2.

## Conflict of Interest

The authors declare no conflict of interest.

## Author Contributions

M.H. and G.S. fabricated and characterized the samples and performed the UV-ARPES measurements and analyzed the ARPES data. M.H. performed the GGA+U DFT calculations and S.W.D and J.M. the DLM-CPA calculations. L.S and J.T. discussed the physics of the altermagnetism and performed the spin-polarized bulk calculations. N.O., T.Z. and G.K. participated in the UV-ARPES measurements and O.C. and J.M. performed the TEM and D.K. the SQUID measurements. All authors discussed the ARPES data and its interpretations. G.S. conceived and supervised the project. M.H. wrote the manuscript with the input from all authors.

## Data Availability Statement

The data that support the findings of this study are available from the corresponding author upon reasonable request.

## References

[1]     I. Žutić, J. Fabian, and S. Das Sarma, "Spintronics: Fundamentals and applications," *Rev. Mod. Phys.*, vol. 76, no. 2, pp. 323–410, 2004, doi: 10.1103/RevModPhys.76.323.

[2]     T. Jungwirth, X. Marti, P. Wadley, and J. Wunderlich, "Antiferromagnetic spintronics," *Nat. Nanotechnol.*, vol. 11, no. 3, pp. 231–241, 2016, doi: 10.1038/nnano.2016.18.

[3]     V. Baltz, A. Manchon, M. Tsoi, T. Moriyama, T. Ono, and Y. Tserkovnyak, "Antiferromagnetic spintronics," *Rev. Mod. Phys.*, vol. 90, no. 1, p. 15005, 2018, doi: 10.1103/RevModPhys.90.015005.

[4]     T. Jungwirth, J. Sinova, A. Manchon, X. Marti, J. Wunderlich, and C. Felser, "The multiple directions of antiferromagnetic spintronics," *Nat. Phys.*, vol. 14, no. 3, pp. 200–203, 2018, doi: 10.1038/s41567-018-0063-6.

[5]     J. Železný, P. Wadley, K. Olejník, A. Hoffmann, and H. Ohno, "Spin transport and spin torque in antiferromagnetic devices," *Nat. Phys.*, vol. 14, no. 3, pp. 220–228, 2018, doi: 10.1038/s41567-018-0062-7.

[6]     L. Šmejkal, J. Sinova, and T. Jungwirth, "Beyond Conventional Ferromagnetism and Antiferromagnetism: A Phase with Nonrelativistic Spin and Crystal Rotation Symmetry," *Phys. Rev. X*, vol. 12, no. 3, pp. 1–16, 2022, doi: 10.1103/PhysRevX.12.031042.

[7]     L. Šmejkal *et al.*, "Crystal time-reversal symmetry breaking and spontaneous Hall effect in collinear antiferromagnets," *Sci. Adv.*, vol. 6, no. 23, 2020, doi: 10.1126/sciadv.aaz8809.

[8]     S. Hayami, Y. Yanagi, and H. Kusunose, "Momentum-dependent spin splitting by collinear antiferromagnetic ordering," *J. Phys. Soc. Japan*, vol. 88, no. 12, 2019, doi: 10.7566/JPSJ.88.123702.

[9]     L. D. Yuan, Z. Wang, J. W. Luo, E. I. Rashba, and A. Zunger, "Giant momentum-dependent spin splitting in centrosymmetric low- Z antiferromagnets," *Phys. Rev. B*, vol. 102, no. 1, p. 14422, 2020, doi: 10.1103/PhysRevB.102.014422.

[10]    R. González-Hernández *et al.*, "Efficient Electrical Spin Splitter Based on Nonrelativistic Collinear Antiferromagnetism," vol. 127701, pp. 1–6, 2021, doi: 10.1103/PhysRevLett.126.127701.

[11]    L. Šmejkal, J. Sinova, and T. Jungwirth, "Emerging Research Landscape of Altermagnetism," *Phys. Rev. X*, vol. 12, no. 4, pp. 1–27, 2022, doi: 10.1103/PhysRevX.12.040501.

[12]    R. D. Gonzalez Betancourt *et al.*, "Spontaneous Anomalous Hall Effect Arising from an Unconventional Compensated Magnetic Phase in a Semiconductor," *Phys. Rev. Lett.*, vol. 130, no. 3, pp. 1–7, 2023, doi: 10.1103/PhysRevLett.130.036702.

[13]    J. Krempaský *et al.*, "Altermagnetic lifting of Kramers spin degeneracy," pp. 1–22, 2023, [Online]. Available: http://arxiv.org/abs/2308.10681

[14]  J. D. Wasscher, "Evidence of weak ferromagnetism in MnTe from galvanomagnetic measurements," *Solid State Commun.*, vol. 3, no. 8, pp. 169–171, 1965, doi: 10.1016/0038-1098(65)90284-X.

[15]  S. Lee *et al.*, "Broken Kramers' degeneracy in altermagnetic MnTe," vol. 036702, 2023, doi: 10.1103/PhysRevLett.132.036702.

[16]  T. Osumi *et al.*, "Observation of Giant Band Splitting in Altermagnetic MnTe," pp. 1–16, 2023, [Online]. Available: http://arxiv.org/abs/2308.10117

[17]  O. Fedchenko *et al.*, "Observation of time-reversal symmetry breaking in the band structure of altermagnetic RuO$_2$," no. Mcd, pp. 1–13, 2023, [Online]. Available: http://arxiv.org/abs/2306.02170

[18]  S. Reimers *et al.*, "Direct observation of altermagnetic band splitting in CrSb thin films," pp. 1–10, 2023, [Online]. Available: http://arxiv.org/abs/2310.17280

[19]  D. Kriegner *et al.*, "Multiple-stable anisotropic magnetoresistance memory in antiferromagnetic MnTe," *Nat. Commun.*, vol. 7, pp. 1–7, 2016, doi: 10.1038/ncomms11623.

[20]  D. Kriegner *et al.*, "Magnetic anisotropy in antiferromagnetic hexagonal MnTe," *Phys. Rev. B*, vol. 96, no. 21, pp. 1–8, 2017, doi: 10.1103/PhysRevB.96.214418.

[21]  D. Bossini *et al.*, "Exchange-mediated magnetic blue-shift of the band-gap energy in the antiferromagnetic semiconductor MnTe," *New J. Phys.*, vol. 22, no. 8, 2020, doi: 10.1088/1367-2630/aba0e7.

[22]  D. Bossini *et al.*, "Femtosecond phononic coupling to both spins and charges in a room-temperature antiferromagnetic semiconductor," *Phys. Rev. B*, vol. 104, no. 22, pp. 1–10, 2021, doi: 10.1103/PhysRevB.104.224424.

[23]  P. E. Faria Junior *et al.*, "Sensitivity of the MnTe valence band to the orientation of magnetic moments," *Phys. Rev. B*, vol. 107, no. 10, pp. 1–5, 2023, doi: 10.1103/PhysRevB.107.L100417.

[24]  G. Yin, J. X. Yu, Y. Liu, R. K. Lake, J. Zang, and K. L. Wang, "Planar Hall Effect in Antiferromagnetic MnTe Thin Films," *Phys. Rev. Lett.*, vol. 122, no. 10, p. 106602, 2019, doi: 10.1103/PhysRevLett.122.106602.

[25]  K. P. Kluczyk *et al.*, "Coexistence of Anomalous Hall Effect and Weak Net Magnetization in Collinear Antiferromagnet MnTe," pp. 1–9.

[26]  T. Komatsubara, M. Murakami, and E. Hirahara, "Magnetic Properties of Manganese Telluride Single Crystals," *J. Phys. Soc. Japan*, vol. 18, no. 3, pp. 356–364, Mar. 1963, doi: 10.1143/JPSJ.18.356.

[27]  P. E. Blöchl, "Projector augmented-wave method," *Phys. Rev. B*, vol. 50, no. 24, pp. 17953–17979, 1994, doi: 10.1103/PhysRevB.50.17953.

[28]  G. Kresse and J. Furthmüller, "Efficiency of ab-initio total energy calculations for metals and semiconductors using a plane-wave basis set," *Comput. Mater. Sci.*, vol. 6, no. 1, pp. 15–50, 1996, doi: 10.1016/0927-0256(96)00008-0.

[29] J. P. Perdew, K. Burke, and M. Ernzerhof, "Generalized gradient approximation made simple," *Phys. Rev. Lett.*, vol. 77, no. 18, pp. 3865–3868, 1996, doi: 10.1103/PhysRevLett.77.3865.

[30] S. Mu *et al.*, "Phonons, magnons, and lattice thermal transport in antiferromagnetic semiconductor MnTe," *Phys. Rev. Mater.*, vol. 3, no. 2, pp. 1–7, 2019, doi: 10.1103/PhysRevMaterials.3.025403.

[31] H. Ebert, A. Perlov, and S. Mankovsky, "Incorporation of the rotationally invariant LDA + U scheme into the SPR-KKR formalism: Application to disordered alloys," *Solid State Commun.*, vol. 127, no. 6, pp. 443–446, 2003, doi: 10.1016/S0038-1098(03)00455-1.

[32] H. Ebert, D. Ködderitzsch, and J. Minár, "Calculating condensed matter properties using the KKR-Green's function method - Recent developments and applications," *Reports Prog. Phys.*, vol. 74, no. 9, 2011, doi: 10.1088/0034-4885/74/9/096501.

[33] P. Lloyd, "Wave propagation through an assembly of spheres: III. The density of states in a liquid," *Proc. Phys. Soc.*, vol. 90, no. 1, pp. 217–231, 1967, doi: 10.1088/0370-1328/90/1/324.

[34] P. Lloyd and P. V. Smith, "Multiple scattering theory in condensed materials," *Adv. Phys.*, vol. 21, no. 89, pp. 69–142, 1972, doi: 10.1080/00018737200101268.

[35] R. Zeller, "Improving the charge density normalization in Korringa-Kohn-Rostoker Green-function calculations," *J. Phys. Condens. Matter*, vol. 20, no. 3, 2008, doi: 10.1088/0953-8984/20/03/035220.

# Supplementary information:

# Temperature dependence of relativistic valence band splitting induced by an altermagnetic phase transition

M. Hajlaoui[1], S.W. D'Souza[2], L. Šmejkal[3,4], D. Kriegner[3], G. Krizman[1], T. Zakusylo[1], N. Olszowska[5], O. Caha[6], J. Michalička[7], A. Marmodoro[2,3], K. Výborný[3], A. Ernst[8], M. Cinchetti[9], J. Minar[2], T. Jungwirth[3,10], and G. Springholz[1]

[1]Institute of Semiconductors and Solid-State Physics, Johannes Kepler University, 4040 Linz, Austria
[2]University of West Bohemia, New Technologies Research Center, Pilsen 30100, Czech Republic
[3]Institute of Physics, Czech Academy of Sciences, Cukrovarnická 10, 162 00 Praha 6, Czech Republic
[4]Institute of Physics, Johannes Gutenberg University Mainz, D-55099 Mainz, Germany
[5]National Synchrotron Radiation Centre SOLARIS, Jagiellonian University, Czerwone Maki 98, 30-392 Krakow, Poland
[6]Department of Condensed Matter Physics, Masaryk University, Kotlářská 267/2, Brno 61137, Czech Republic
[7]Central European Institute of Technology, Brno University of Technology, Purkyňova 123, Brno, 61200, Czech Republic
[8]Institute for Theoretical Physics, Johannes Kepler University, 4040 Linz, Austria
[9]Department of Physics, TU Dortmund University, 44227 Dortmund, Germany
[10]School of Physics and Astronomy, University of Nottingham, Nottingham NG7 2RD, United Kingdom

**SUPPLEMENTRAY NOTE1**: ARPES versus GGA + U with and without SOC

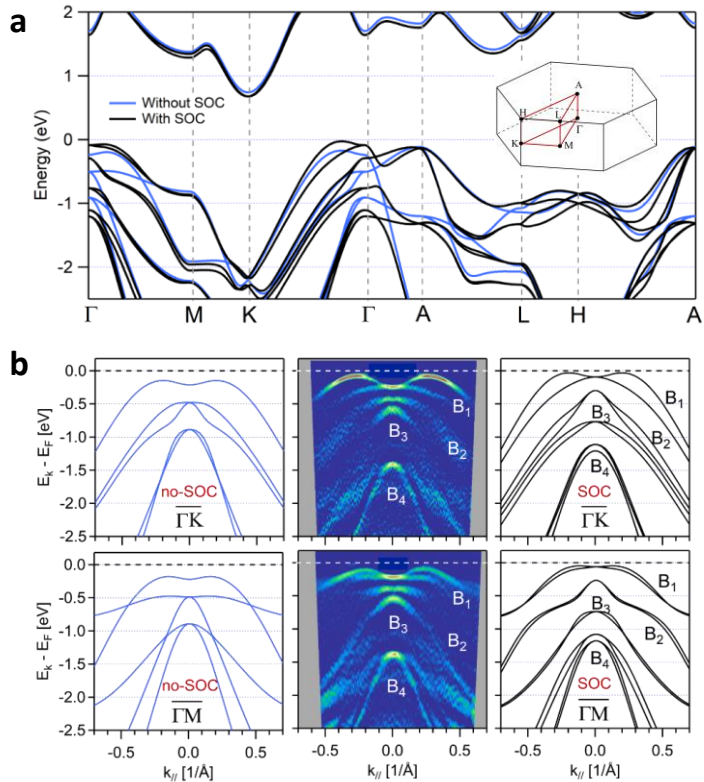

**Figure S1:** Electronic structure of α-MnTe along different high symmetry directions with and without spin-orbit coupling (SOC) included. (a) Full band dispersions along all high symmetry directions. (b) Comparison between the GGA+U calculations with the ARPES data (middle) without SOC (left hand side) and with SOC (right hand side) for both the ΓK (top) and ΓM (bottom) directions.

To illustrate the relativistic effect on the valence band splitting, we present in Figure S1(a) the GGA+U calculation with and without SOC along different high symmetry directions. In Figure S1(b) we show the comparison between the ARPES data at hv = 21 eV with the calculation, with and without SOC. The splitting of bands due to SOC effect is clearly evident in the ARPES images.

## SUPPLEMENTRAY NOTE2: Impact of Spin-orbit coupling on the in-plane electronic structure with a Néel vector orientation along the [1$\bar{1}$00] direction

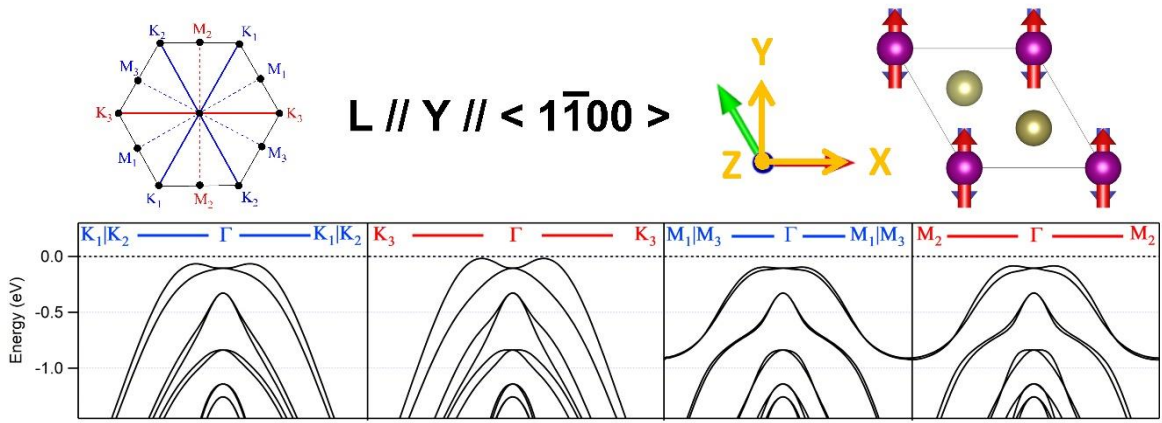

**Figure S2:** Electronic structure of α-MnTe with spin-orbit coupling, along the different in-plane direction, $\Gamma K_i$ and $\Gamma M_i$, in the presence of Néel vector along the [1$\bar{1}$00] direction.

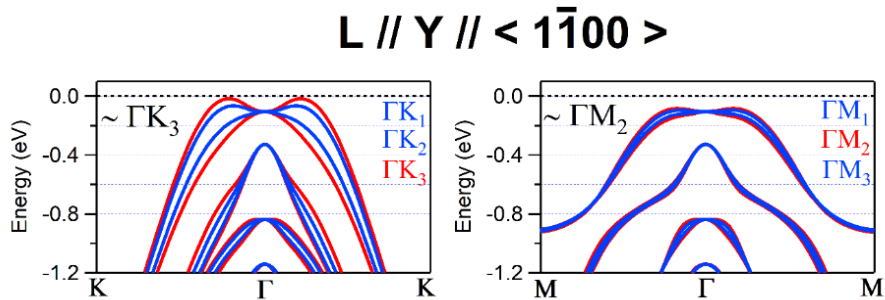

**Figure S3**: **The effect of domain averaging on the electronic structure with SOC in case of Néel vector along the [1$\bar{1}$00] direction**. On the left, the superimposed electronic structure near the Fermi level along the ΓK directions averaging over three equivalent magnetic domains *i* = 1, 2 and 3 as shown in Fig. S1. For each ΓK direction, the averaging results in the superposition of two equivalents $\Gamma K_1$ and $\Gamma K_2$ electronic structures (in blue), and a distinct $\Gamma K_3$ electronic structure (in red). On the right, the superimposed electronic structure is shown for the ΓM direction after averaging over the same three equivalent magnetic domains. Here, each ΓM direction combines two equivalents $\Gamma M_1$ and $\Gamma M_3$ set of bands (in blue) and one $\Gamma M_2$ set of bands (in red). Due to the smaller spin splitting along the $\Gamma M_i$ directions, the splitting in the superimposed band structure is difficult to discern.

The interplay of magnetic order with spin-orbit coupling significantly influences both the material's symmetry and its electronic structure. As discussed in Ref. **[23],** in the paramagnetic phase, α-MnTe possesses the D6h point group symmetry. When magnetic order is present, this symmetry is reduced to its subgroup D3d, when spin-orbit is neglected. The orientation of the Néel vector results in two distinct scenarios when spin-orbit coupling is present:

(i) In the case of an in-plane Néel vector along the $[1\bar{1}00]$, the D3d group symmetry further reduces to C2h group symmetry, introducing an in-plane electronic structure anisotropy. This effect is illustrated in Fig. S1, demonstrating an in-plane two-fold symmetry (red and blue spectra).

(ii) Conversely, when the Néel vector points out of the plane, along the [0001] direction, the D3d group symmetry remains unchanged, resulting in in-plane electronic structure with six-fold symmetry.

In Fig. S2, we offer insights into the impact of domain averaging on the anisotropic electronic structure. Our results illustrate that averaging over the three magnetic domains leads to only moderate variations in the band. Notably, we find that the band structure along each ΓK (ΓM) direction can be similar to the results obtained for the case of the $\Gamma K_3$ ($\Gamma M_2$) direction.

## SUPPLEMENTRAY NOTE3: Temperature dependence of UV-ARPES spectra at h$\nu$ = 21 eV and h$\nu$ = 48 eV: raw data and derivatives

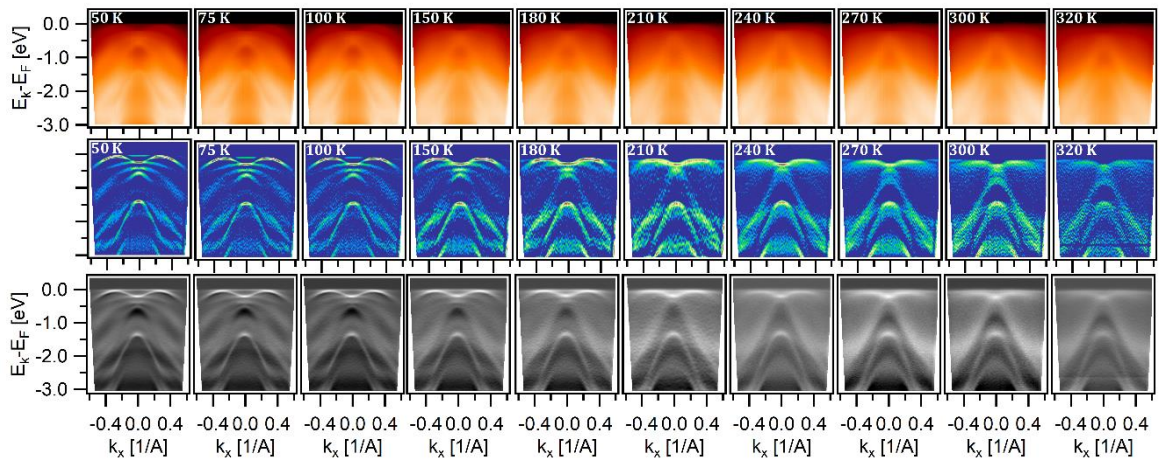

**Figure S4: Temperature dependence of the ARPES spectra along ΓK recorded at h$\nu$ = 21 eV.** Top panel: Raw data, middle panel: second derivative, and bottom panel: first derivative.

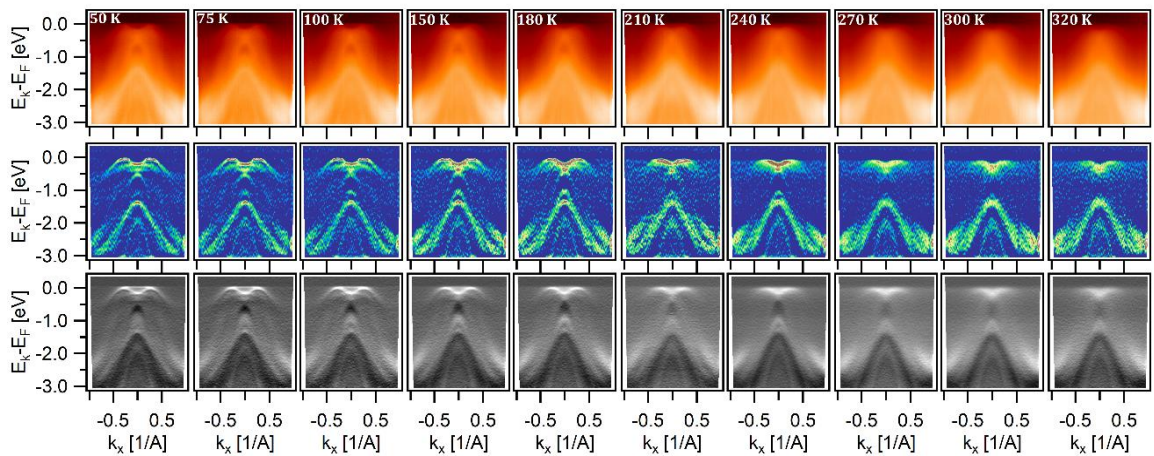

**Figure S5: Temperature dependence of the ARPES spectra along ΓK recorded at h$\nu$ = 48 eV.** Top panel: Raw data, middle panel: second derivative, and bottom panel: first derivative.

15